# A note on a universal random variate generator for integer-valued random variables

L. Barabesi[1] and L. Pratelli[2]
[1]Dipartimento di Economia Politica, Università di Siena,
P.zza S.Francesco 17, 53100 Siena, Italy
[2]Accademia Navale, viale Italia 72, 57100 Livorno, Italy

**Abstract.** A universal generator for integer-valued square-integrable random variables is introduced. The generator relies on a rejection technique based on a generalization of the inversion formula for integer-valued random variables. The proposal gives rise to a simple algorithm which may be implemented in a few code lines and which may show good performance when the classical families of distributions - such as the Poisson and the Binomial - are considered. In addition, the method is suitable for the computer generation of integer-valued random variables which display closed-form characteristic functions, but do not possess a probability function expressible in a simple analytical way. As an example of such a framework, an application to the Poisson-Tweedie distribution is provided.

**1. Introduction.** As emphasized by Johnson *et al.* (2005), discrete distributions commonly adopted in statistics are members of the lattice distribution family. In turn, by using a straightforward linear transformation, it is generally convenient to consider random variables defined on the integers rather than on a lattice. Hence, the computer generation of integer-valued random variables is obviously of central importance for simulation purposes. In such a case, the inversion, the alias and the rejection methods usually constitute the general tools adopted to carry out suitable algorithms. For an extensive treatment of the aforementioned methods, see Devroye (1986, p.83-116) and Hörmann *et al.* (2004, p.43-52).

One of the recent research trend for nonuniform random variate generation is focused on the implementation of algorithms which are suitable for large families of distributions. These algorithms are usually referred to as universal (also automatic or black-box) generators (Devroye, 1986, p.286-356). For a monograph on the topic, where the advantages of this kind of generators are extensively described, see Hörmann *et al.* (2004). Universal generators have been mainly explored when absolutely continuous random variables are considered. As to integer-valued random variables, the inversion and the alias methods may actually give rise to universal generators. However, the inversion method generally produces quite inefficient algorithms and the alias method is suited for random variables taking values on a finite subset of integers. In order to achieve methods displaying bounded expected complexity, Devroye (1987) and Hörmann (1994) propose universal algorithms based on rejection methods for integer-valued log-concave distributions (see also Devroye, 2012). In addition, Leydold (2001, 2003) analyzes similar techniques for the wider family of integer-valued T-concave distributions. It should be remarked that these methods require the knowledge of the mode of the distribution, which may be not an easy task to accomplish in some cases. Therefore, even if the log-concave and T-concave integer-valued distribution families encompass most of the classical distributions, a method requiring milder assumptions could be useful.

The present paper aims to introduce a universal generator for integer-valued square-integrable random variables. The algorithm stems on a simple generalization of the inversion formula for integer-valued random variables, which does not seem previously given in statistical literature. This formula gives rise to a suitable inequality, which could be immediately adopted for implementing a generator for a given specific distribution. However, since our target is focused on the achievement of a universal generator, we suitably consider a special case of the inequality in order to provide a rejection method - which actually constitutes the counterpart for integer-valued random variables of the Devroye's (1981) proposal. More precisely, we introduce a suitable dominating function which resembles the hat function usually adopted in the ratio-of-uniforms method. The corresponding dominating probability function is characterized by coefficients depending on the characteristic function of the integer-valued random variable to be generated. The resulting algorithm may be easily implemented in few code lines (with a small set-up) and may be appealing for the generation of random variables possessing a symmetric (or nearly symmetric) unimodal probability function. In addition, the algorithm could be very suitable when the integer-valued random variable solely shows a closed-form characteristic function (for the equivalent problem in the case of an absolutely continuous random variable, see Devroye, 1981, Ridout, 2009).

The present paper has been motivated by the practical implementation of a random variate generator for the Poisson-Tweedie distribution. This distribution family is very flexible and encompasses as special cases many commonly-used distributions such as the Poisson, the Negative Binomial, the Poisson Inverse Gaussian, the Discrete Stable, the Poisson-Pascal and the Neyman Type A (see El-Shaarawi *et al.*, 2009, Johnson *et al.,* 2005). Moreover, the Poisson-Tweedie distribution is very useful for modeling overdispersed count data in biological and environmental studies (Hougaard *et al.*, 1997, El-Shaarawi *et al.*, 2009). Regrettably, the probability function of the Poisson-Tweedie random variable is not known in a simple form and its computation is implemented by means of recurrence formulas (El-Shaarawi *et al.*, 2009). Accordingly, the suggested algorithm seems to be very suitable when simulation studies involving the Poisson-Tweedie distribution are carried out.

The outline of the paper is as follows. In Section 2, the main results are introduced and the suggested algorithm is presented. Section 4 shows some illustrative applications of the proposal to the Poisson and Binomial distributions. In Section 4, the random variate generation for the Poisson-Tweedie distribution is considered. Finally, in Section 5 the conclusions are drawn and some suggestions for the future research are given.

**2. The proposed method.** Let $X$ be an integer-valued random variable (r.v.) with probability function (p.f.) given by $p_X(x) = P(X = x)$ and characteristic function (c.f.) given by $\phi_X(t) = \mathrm{E}[e^{itX}]$, where - as usual - the symbol $i$ represents the imaginary unit. First, we provide a useful integral representation for $p_X$ in the following result.

**Result 1.** Let $X$ be an integer-valued r.v. with p.f. $p_X$. Moreover, let $g$ be a measurable function defined on $\mathbb{Z}$, such that $\mathrm{E}[|g(X)|] < \infty$. If $x \in \mathbb{Z}$ and $g(x) \neq 0$, it holds

$$p_X(x) = \frac{1}{2\pi g(x)} \int_{-\pi}^{\pi} e^{-itx} \mathrm{E}[g(X)e^{itX}] \, \mathrm{d}t \ .$$

**Proof.** The expectation $\mathrm{E}[g(X)e^{itX}]$ exists, since from the assumptions it follows that

$$|\mathrm{E}[g(X)e^{itX}]| \leq \mathrm{E}[|g(X)e^{itX}|] = \mathrm{E}[|g(X)|] < \infty \,.$$

Hence, it holds

$$\frac{1}{2\pi} \int_{-\pi}^{\pi} e^{-itx} \mathrm{E}[g(X)e^{itX}] \, \mathrm{d}t = \frac{1}{2\pi} \int_{-\pi}^{\pi} \sum_{y \in \mathbb{Z}} g(y) e^{it(y-x)} p_X(y) \, \mathrm{d}t$$

$$= \frac{1}{2\pi} \sum_{y \in \mathbb{Z}} g(y) p_X(y) \int_{-\pi}^{\pi} e^{it(y-x)} \, \mathrm{d}t = g(x) p_X(x) \,,$$

owing to the orthogonality of the functions $(t \mapsto e^{itn})_{n \in \mathbb{Z}}$. $\square$

It should be noticed that Result 1 actually provides a generalization of the well-known inversion formula (see *e.g.* Feller, 1971, p.511)

$$p_X(x) = \frac{1}{2\pi} \int_{-\pi}^{\pi} e^{-itx} \phi_X(t) \, \mathrm{d}t \,, \tag{1}$$

which is achieved by applying Result 1 with $g \equiv 1$. In turn, from Result 1, it easily follows the inequality

$$p_X(x) \leq \frac{1}{2\pi |g(x)|} \int_{-\pi}^{\pi} |\mathrm{E}[g(X)e^{itX}]| \, \mathrm{d}t = \frac{1}{\pi |g(x)|} \int_{0}^{\pi} |\mathrm{E}[g(X)e^{itX}]| \, \mathrm{d}t \,, \tag{2}$$

where the last identity is obtained by noticing that $|\mathrm{E}[g(X)e^{itX}]| = |\mathrm{E}[g(X)e^{-itX}]|$. Inequality (2) could be immediately adopted in order to implement a rejection algorithm by choosing $g(x)^{-1}$ as the p.f. of an *ad hoc* integer-valued r.v. However, since the aim of the present paper is to provide an easy-to-implement universal generator, rather than a generator for a specific r.v., suitable selections for $g$ are carried out in the spirit of the algorithm given by Devroye (1981) for absolutely continuous r.v.'s.

If $g \equiv 1$, inequality (2) reduces to

$$p_X(x) \leq \frac{1}{\pi} \int_{0}^{\pi} |\phi_X(t)| \, \mathrm{d}t \,.$$

Moreover, let us consider the function $g(x) = (x - m)^2$, in such a way that (1) gives rise to

$$p_X(x) \leq \frac{1}{\pi(x-m)^2} \int_{0}^{\pi} |\mathrm{E}[(X-m)^2 e^{itX}]| \, \mathrm{d}t \,.$$

Hence, by assuming that $m \in \mathbb{Z}$, let us introduce the integer-valued r.v. $Y = X - m$. Since

$$\phi_Y(t) = \exp(-itm) \phi_X(t) \,,$$

the previous inequalities respectively reduce to

$$p_X(x) \leq \frac{1}{\pi} \int_{0}^{\pi} |\phi_Y(t)| \, \mathrm{d}t = \frac{1}{\pi} \int_{0}^{\pi} |\phi_X(t)| \, \mathrm{d}t$$

and, if $\mathrm{E}[X^2] < \infty$,

$$p_X(x) \leq \frac{1}{\pi(x-m)^2} \int_{0}^{\pi} |\mathrm{E}[Y^2 e^{itY}]| \, \mathrm{d}t = \frac{1}{\pi(x-m)^2} \int_{0}^{\pi} |\phi_Y''(t)| \, \mathrm{d}t \,.$$

Thus, it finally follows the inequality

$$p_X(x) \leq \min(c, k_m(x-m)^{-2}),  \quad (3)$$

where

$$c = \frac{1}{\pi} \int_0^\pi |\phi_Y(t)| \, dt$$

and

$$k_m = \frac{1}{\pi} \int_0^\pi |\phi_Y''(t)| \, dt .$$

It is worth remarking that, since the integration of bounded functions is carried out over a finite range, no numerical problems should be encountered in the practical computation of the constants $c$ and $k_m$.

Inequality (3) cannot be immediately adopted for implementing a hybrid rejection method giving rise to a ratio-of-uniforms algorithm (similarly to the method given by Devroye, 1981, for absolutely continuous r.v.'s), since this inequality does not provide a dominating p.f. when truncation (or rounding) is applied. For more details on the hybrid rejection method, see Devroye (1986, p.115). Hence, in order to carry out a suitable rejection method in this case, it is convenient to consider the absolutely continuous r.v. $V$ with probability density function (p.d.f.) given by

$$f_V(v; m, \sigma, \alpha) = \frac{\alpha}{2\sigma} I(|v-m| \leq \sigma) + \frac{(1-\alpha)\sigma}{2(v-m)^2} I(|v-m| > \sigma),$$

where $\alpha \in (0,1)$ and $\sigma > 0$, while $I$ represents the usual indicator function. Hence, $f_V$ is a mixture of two simple p.d.f.'s and the random generation of the r.v. $V$ is straightforward. Actually, $f_V$ is the generalization of the p.d.f. usually adopted for implementing the ratio-of-uniforms method, which is achieved for $\alpha = 1/2$. Indeed, if $U_1$ and $U_2$ represent two independent r.v.'s uniformly distributed on $[0,1]$ and $[-1,1]$ respectively, it turns out that

$$\frac{V-m}{\sigma} \stackrel{\mathcal{L}}{=} I(U_1 \leq \alpha) U_2 + I(U_1 > \alpha) \frac{1}{U_2} .$$

Moreover, if $m \in \mathbb{Z}$ and $j \in \mathbb{N}$ in such a way that $\sigma = j + 1/2$, for $z \in \mathbb{Z}$ the p.f. of the integer-valued r.v. $Z = \text{Round}(V)$ is given by

$$p_Z(z; m, \sigma, \alpha) = \frac{\alpha}{2\sigma} I(|z-m| \leq \sigma) + \frac{(1-\alpha)\sigma}{2(z-m)^2 - 1/2} I(|z-m| > \sigma) .$$

Accordingly, the following result provides the key inequality for the generation of the integer-valued r.v. $X$ by means of the proposed rejection method.

**Result 2.** Let $X$ be an integer-valued r.v. with p.f. $p_X$. If $E[X^2] < \infty$, for $x \in \mathbb{Z}$ it holds

$$p_X(x) \leq A_m p_Z(x; m, \sigma_m, \alpha_m),$$

where $m \in \mathbb{Z}$, $\sigma_m = \text{Round}(\sqrt{k_m/c}) + 1/2$, $\alpha_m = 2\sigma_m c / A_m$ and $A_m = 2(\sigma_m c + k_m/\sigma_m)$.

**Proof.** First, let us assume that $|x - m| \leq \sigma_m$. Hence, it holds

$$c = \frac{\alpha_m A_m}{2\sigma_m} = A_m p_Z(x; m, \sigma_m, \alpha_m) \,.$$

In turn, if $|x - m| > \sigma_m$, it follows that

$$\frac{k_m}{(x-m)^2} = \frac{(1-\alpha_m)A_m \sigma_m}{2(x-m)^2} \leq \frac{(1-\alpha_m)A_m \sigma_m}{2(x-m)^2 - 1/2} = A_m p_Z(x; m, \sigma_m, \alpha_m)$$

and the required inequality follows from (3). □

On the basis of Result 2, a simple rejection algorithm may be easily implemented for the random generation of the integer-valued r.v. $X$. More precisely, by setting

$$h(x; m, c, k_m, \sigma_m) = cI(|x-m| \leq \sigma_m) + \frac{k_m}{(x-m)^2 - 1/4} I(|x-m| > \sigma_m) \,,$$

the following algorithm may be considered.

```
compute  m, c, k_m, σ_m, α_m
repeat
     generate U_1, U_3 uniformly on [0,1]
     generate U_2 uniformly on [−1,1]
     if U_1 > α_m set U_2 := 1/U_2
     set X := Round(m + σ_m U_2)
until p_X(X) < h(X; m, c, k_m, σ_m) U_3
return X
```

In the previous algorithm, when solely $\phi_X$ is known in a closed form, $p_X$ is computed by means of the inversion formula (1). In this case, even if the algorithm may result less efficient in terms of computation timing, no numerical drawbacks should occur when an appropriate software is adopted, since the numerical integration of a bounded function is carried out over a finite range. In addition, it is worth noting that the expected complexity - *i.e.* the expected number of iterations in the algorithm - is given by $A_m$ (see the inequality provided in Result 2). Obviously, given the morphology of $p_Z$, the algorithm should be particularly efficient for random generation of a r.v. $X$ displaying a symmetric unimodal p.f.

For the sake of simplicity, the algorithm is given in its very basic version, even if it could be suitably improved by means of an appropriate handling of the function $h$ and by means of some squeezing pre-tests. Moreover, from the definition of $\sigma_m$, it should be apparent that the value of $\alpha$ is nearly equal to $1/2$ and hence the method may be actually considered as the counterpart for integer-valued r.v.'s of the Devroye's (1981) technique tailored for absolutely continuous r.v.'s. Finally, it should be remarked that the algorithm could be very appealing for the generation of a sum of *iid* r.v.'s, since - as is well known - the c.f. of the sum displays a manageable form.

The final remark obviously deals with the appropriate selection of $m$. To this aim, a natural choice of $m$ is given by the value which minimizes the expected complexity $A_m$. However, the minimization step may be time-consuming at the set-up stage and hence alternative selections could be considered. Owing to the choice of $\sigma_m$, the expected complexity should be nearly equivalent to $4\sqrt{ck_m}$ in practice. Hence, since $c$ does not depend on $m$, a sub-optimal selection of $m$ should be given by $m^* = \text{Round}(\arg\min k_m)$. In any case, it should be remarked that this choice is optimal for achieving the tightness of inequality (3). Alternatively, since

$$|\phi_Y''(t)| \leq \mathrm{E}[(X-m)^2]$$

and hence

$$k_m \leq \mathrm{E}[(X-m)^2] \,,$$

the selection $m^{**} = \mathrm{Round}(\mathrm{E}[X])$ might be considered in order to obtain an even simpler set-up stage. Indeed, $m^{**}$ is easily computed by taking into account that $\mathrm{E}[X] = -i\phi_X'(0)$.

**3. Application to the Poisson and Binomial distributions.** In order to assess the efficiency of the proposed algorithm, some classical families of integer-valued r.v.'s are considered. First, let us assume that $X$ be a Poisson r.v. with parameter $\lambda \in (0, \infty)$, in such a way that - as is well known - the corresponding c.f. is given by

$$\phi_X(t) = e^{\lambda(e^{it}-1)} \,.$$

Even if there exists very efficient algorithms for the random generation of Poisson r.v.'s (see *e.g.* Devroye, 1986), Ahrens and Dieter (1991) and Stadlober (1990) advocate for the ratio-of-uniforms method as a simple and effective technique to be adopted in this setting (see also Hörmann, 1994). Their proposal, specifically tailored for the Poisson distribution, is rather similar in its concept to the universal method presented in the previous section and hence a performance comparison of the two methods may be worthwhile. In this case, the values of $A_m$ for $m = m^*$ and $m = m^{**} = \mathrm{Round}(\lambda)$ have been computed for various choices of $\lambda$. For all the considered values of $\lambda$, it was found that $m^* = m^{**}$. In turn, the expected complexity (say $A_{\mathrm{AD}}$) of the algorithm proposed by Ahrens and Dieter (1991), and optimized as suggested by Stadlober (1990), has been computed for the same values of $\lambda$. The corresponding results are reported in Table I.

**Table I.** Expected complexities $A_{m^*}$ and $A_{\mathrm{AD}}$ for the considered algorithms (Poisson distribution).

| $\lambda$ | 1 | 2 | 5 | 10 | 20 | 50 | 100 | $\infty$ |
|---|---|---|---|---|---|---|---|---|
| $A_{m^*}$ | 1.99 | 1.83 | 1.66 | 1.61 | 1.59 | 1.58 | 1.58 | 1.57 |
| $A_{\mathrm{AD}}$ | 2.21 | 1.91 | 1.70 | 1.60 | 1.53 | 1.46 | 1.43 | 1.37 |

As it can be assessed from Table I, the two algorithms produce quite similar performance, even if the Ahrens and Dieter (1991) method is expressly implemented for the Poisson distribution. In addition, the expected complexity of the proposed algorithm converges very fast to the limit. Indeed, if the ratio-of-uniforms algorithm proposed by Devroye (1981) is considered for random generation from the Normal distribution, the expected complexity is given by $(512/(\mathrm{e}\pi^3))^{1/4} \simeq 1.57$.

In the second study, a Binomial r.v. $X$ with parameters $n \in \mathbb{N}$ and $p \in (0,1)$ is assumed. Obviously, the corresponding c.f. is given by

$$\phi_X(t) = (1-p+pe^{it})^n \,.$$

In turn, the values of $A_m$ for $m = m^*$ and $m = m^{**} = \mathrm{Round}(np)$ have been computed for various choices of $n$ and $p$. For all the considered values of $\lambda$, it was found that $m^* = m^{**}$. Moreover, the expected complexity (say $A_{\mathrm{S}}$) of the ratio-of-uniforms algorithm proposed by Stadlober (1989, 1990), an easy ratio-of-uniforms algorithm for the random generation of

Binomial r.v.'s (see also Johnson *et al.*, 2005, p.126), has been computed for the same parameter values and the results are reported in Table II.

**Table II.** Expected complexities $A_{m^*}$ and $A_S$ for the considered algorithms (Binomial distribution).

|   |     |           | 10   | 20   | 40   | 100  | 200  | 400  | $\infty$ |
|---|-----|-----------|------|------|------|------|------|------|----------|
|   |     |  $n$      |      |      |      |      |      |      |          |
| $p$ | 0.1 | $A_{m^*}$ | 1.94 | 1.77 | 1.71 | 1.62 | 1.59 | 1.58 | 1.57 |
|   |     | $A_S$     | 2.21 | 1.86 | 1.71 | 1.60 | 1.52 | 1.48 | 1.37 |
|   | 0.2 | $A_{m^*}$ | 1.72 | 1.71 | 1.67 | 1.59 | 1.59 | 1.58 | 1.57 |
|   |     | $A_S$     | 1.80 | 1.73 | 1.62 | 1.52 | 1.47 | 1.44 | 1.37 |
|   | 0.3 | $A_{m^*}$ | 1.61 | 1.61 | 1.61 | 1.60 | 1.57 | 1.57 | 1.57 |
|   |     | $A_S$     | 1.80 | 1.64 | 1.57 | 1.49 | 1.46 | 1.43 | 1.37 |
|   | 0.4 | $A_{m^*}$ | 1.75 | 1.59 | 1.58 | 1.58 | 1.58 | 1.58 | 1.57 |
|   |     | $A_S$     | 1.74 | 1.62 | 1.54 | 1.47 | 1.44 | 1.42 | 1.37 |
|   | 0.5 | $A_{m^*}$ | 1.73 | 1.58 | 1.58 | 1.58 | 1.57 | 1.57 | 1.57 |
|   |     | $A_S$     | 1.70 | 1.60 | 1.52 | 1.47 | 1.44 | 1.42 | 1.37 |

By inspecting Table II, it is at once apparent that the same comments introduced for the Poisson distribution study also apply to the Binomial distribution case. In addition, similar patterns have been assessed for the Hypergeometric distribution in comparison with the ratio-of-uniforms approach proposed by Stadlober (1990), even if the results are not reported. Finally, it should be emphasized that the proposed universal algorithm often performs better than the *ad hoc* ratio-of-uniforms algorithms as shown in Tables I and II, since its implementation is based on a slight different background. Indeed, the dominating p.f. is achieved by means of the rounding of a r.v., rather than a truncation of a r.v. (as in the case of the proposals by Ahrens and Dieter, 1991, and by Stadlober, 1990). This issue gives rise to a "more centered" dominating p.f., which in turn may provide a smaller expected complexity.

**4. Application to the Poisson-Tweedie distribution.** An integer-valued r.v. $X$ with the Poisson-Tweedie distribution displays the c.f.

$$\phi_X(t) = e^{(b/a)[(1-c)^a - (1-ce^{it})^a]} ,$$

where $(a, b, c) \in \{(-\infty, 0] \times (0, \infty) \times [0, 1)\} \cup \{(0, 1] \times (0, \infty) \times [0, 1]\}$ (see Johnson *et al.*, 2005, p.480). It should be remarked that the parameterization suggested by El-Shaarawi *et al.* (2009) is adopted in the definition of $\phi_X$.

The generation of a random variate from the Poisson-Tweedie distribution is not difficult when $a \in (-\infty, 0)$, since in this case $\phi_X$ may be rewritten as

$$\phi_X(t) = e^{-(b/a)(1-c)^a[\phi_Y(t)-1]} ,$$

where

$$\phi_Y(t) = \left(\frac{1-c}{1-ce^{it}}\right)^{-a}$$

represents the c.f. of a Negative Binomial r.v. $Y$ with parameters $(-a)$ and $(1-c)$ (we readopt the symbol $Y$ with a slight abuse in notation). Hence, the r.v. $X$ may be represented as a compound Poisson r.v. and, since the sum of *iid* Negative Binomial r.v.'s is in turn a Negative Binomial r.v., the random generation of the r.v. $X$ is straightforward. Obviously, the generation is even more immediate for $a=0$, since in this case the Poisson-Tweedie distribution reduces to a Negative Binomial distribution. In contrast, when $a \in (0,1]$, the c.f. $\phi_X$ may be reformulated as

$$\phi_X(t) = e^{(b/a)[\phi_Y(t)-1]},$$

where, in this case, the r.v. $Y$ represents a down-weighted Sibuya r.v. with c.f.

$$\phi_Y(t) = (1-c)^a + 1 - (1-ce^{it})^a$$

(see also Zhu and Joe, 2009). Thus, the r.v. $X$ may be again represented as a compound Poisson r.v., even if the sum of *iid* down-weighted Sibuya r.v.'s cannot be easily managed. Therefore, in this case the representation is not useful for achieving a suitable generator.

It is worth noting that the Poisson-Tweedie distribution may be also seen as a mixture Poisson distribution, with a mixturing absolutely continuous Tweedie r.v. (see Aalen, 1992, Hougaard *et al.*, 1997). This stochastic representation permits for an alternative random variate generation for the Poisson-Tweedie distribution. However, while it is straightforward to generate an absolutely continuous Tweedie random variate when $a \in (-\infty, 0]$ - since it solely involves a Gamma random variate - the task is not trivial when $a \in (0,1]$ - since in this case it requires an exponentially-tilted Stable random variate, which is quite difficult to generate (Devroye, 2009). Incidentally, it should be also remarked that, when $a \in (0,1]$, it holds

$$\phi_X(t) = e^{(b/a)(1-c)^a} e^{-(b/a)(1-ce^{it})^a}$$
$$= e^{(b/a)(1-c)^a} \sum_{x \in \mathbb{N}} e^{itx} c^x p_Y(x),$$

where, with the usual slight abuse in notation, in this case the r.v. $Y$ represents a Discrete Stable r.v. with parameters $a$ and $b/a$ and $p_Y$ is the corresponding p.f. (for more details on the Discrete Stable r.v., see *e.g.* Marcheselli *et al.*, 2008). Hence, from the definition of c.f. it readily follows that

$$p_X(x) = e^{(b/a)(1-c)^a} c^x p_Y(x).$$

Thus, the r.v. $X$ may be actually considered as an exponentially-tilted Discrete Stable r.v., *i.e.* the integer-valued counterpart of an exponentially-tilted Stable r.v. However, this stochastic representation does not lead to a finite-time algorithm, except than for the special case $c=1$, *i.e.* when the Poisson-Tweedie distribution reduces to the Discrete Stable distribution and random variate generation may be accomplished by means of the proposal by Devroye (1993).

On the basis of the previous discussion, it is apparent that random variate generation for the Poisson-Tweedie distribution may be challenging when $a \in (0,1]$ and $c \neq 1$. In this case, the suggested universal algorithm could be suitable, also by taking into account that the p.f. of the r.v. $X$ is unimodal for this parameter range (see Hougaard *et al.*, 1997). Since $E[X^2] < \infty$ if $c \neq 1$, the values of $A_m$ for $m = m^*$ and $m = m^{**} = \text{Round}(bc/(1-c)^{1-a})$ have been computed for various choices of $a$, $b$ and $c$ and the results are reported in Table III. For some parameter choices it occurred that $m^* \neq m^{**}$ and hence the expected complexities $A_{m^*}$ and $A_{m^{**}}$ were both reported in Table III. From an analysis of Table III, it is apparent that the algorithm performs reasonably well, even if it is suitable to adopt the selection $m^*$, which in some cases may markedly lessen the expected complexity.

**Table III.** Expected complexities $A_{m^*}$ and $A_{m^{**}}$ (in parenthesis) for the proposed algorithm (Poisson–Tweedie distribution).

| | | | | $c$ | 0.1 | 0.3 | 0.5 | 0.7 | 0.9 |
|---|---|---|---|---|---|---|---|---|---|
| $b$ | 1 | $a$ | 0.1 | | 1.28(1.28) | 2.54(2.54) | 2.44(2.44) | 2.71(3.08) | 3.32(4.73) |
| | | | 0.3 | | 1.27(1.27) | 2.55(2.55) | 2.49(2.49) | 2.32(3.32) | 3.03(4.71) |
| | | | 0.5 | | 1.27(1.27) | 2.56(2.56) | 2.41(2.56) | 2.21(2.21) | 2.64(3.74) |
| | | | 0.7 | | 1.27(1.27) | 2.58(2.58) | 2.41(2.63) | 2.21(2.21) | 2.16(3.07) |
| | | | 0.9 | | 1.27(1.27) | 2.58(2.58) | 2.42(2.71) | 2.30(2.30) | 2.02(2.02) |
| $b$ | 5 | $a$ | 0.1 | | 2.43(2.69) | 2.08(2.08) | 1.97(2.11) | 1.92(1.93) | 1.91(1.97) |
| | | | 0.3 | | 2.43(2.70) | 2.27(2.27) | 2.03(2.03) | 1.90(1.95) | 2.02(2.15) |
| | | | 0.5 | | 2.43(2.71) | 2.35(2.35) | 1.90(2.23) | 1.93(1.93) | 2.06(2.15) |
| | | | 0.7 | | 2.44(2.73) | 2.00(2.44) | 1.90(1.90) | 1.89(1.89) | 1.98(2.10) |
| | | | 0.9 | | 2.44(2.74) | 1.95(2.54) | 1.98(2.04) | 1.78(1.78) | 1.78(1.94) |

**5. Conclusions and future directions.** When the random generation of a square-integrable integer-valued random variable is required, the considered universal method may be suitable. Indeed, the suggested algorithm may be easily implemented in a few code lines with a rather small set-up. In addition, the algorithm shows a good performance in terms of expected complexity for the classical distributions, such as the Poisson, the Binomial and the Hypergeometric distributions. Moreover, the method is potentially useful for the generation of an integer-valued random variable solely displaying closed-form characteristic function - a feature of many important distribution families, see Johnson *et al.* (2005).

As to the future research, it could be interesting to explore the use of inequality (3) for implementing more specialized rejection algorithms, giving rise to fast generators for specific distributions. Actually, the present paper aims to provide a universal generator - with a structure similar to the proposal by Devroye (1981) in the absolutely continuous case - and hence the issue should be explored in a more detail. Moreover, the suggested method could be generalized by adopting the two-parameter choice of type $g(x) = (x - m)^\gamma$, where $\gamma > 0$, in inequality (3). Actually, we have opted for the selection $\gamma = 2$ in order to achieve a very easy-to-implement method. However, the two-parameter choice should lead to a more efficient algorithm when $\gamma$ is properly selected. In addition, random variate generation for heavy-tailed distributions could be handled in this case and the assumption of square-integrability on the underlying random variable removed.